\documentclass[epj]{svjour}
\usepackage{amsfonts}
\usepackage{amsmath}
\usepackage{graphicx}
\usepackage{subfigure}
\usepackage{dcolumn}
\usepackage{bm}
\usepackage{booktabs}
\usepackage{color}

\newcommand{\figcaption}{\def\@captype{figure}\caption}
\newcommand{\tabcaption}{\def\@captype{table}\caption}

\newcommand{\Rmnum}[1]{\expandafter\@slowromancap\romannumeral #1@}
\def\hlinewd#1{%
  \noalign{\ifnum0=`}\fi\hrule \@height #1 \futurelet
   \reserved@a\@xhline}
\makeatother

\def\qq{\langle\bar qq\rangle}

\def\GGb{\langle g_s^2GG\rangle}

\def\qGqa{\langle\bar  qGq\rangle}
\def\qGqb{\langle\bar qg_s\sigma\cdot Gq\rangle}

\def\f(s){\left[(\alpha+\beta)m_c^2-\alpha\beta s\right]}
\def\non{\\ \nonumber}

\usepackage{ulem}

\usepackage{graphics}
\begin{document}

\title{$Z_c(4200)^+$ decay width as a charmonium-like tetraquark state}

\author{Wei Chen\inst{1} \and T. G. Steele\inst{1} \and Hua-Xing Chen\inst{2} \and Shi-Lin Zhu\inst{3,4,5}
}                     % Do not remove
\offprints{}          % Insert a name or remove this line
\institute{Department of Physics
and Engineering Physics, University of Saskatchewan, Saskatoon, SK, S7N 5E2, Canada
\and
School of Physics and Nuclear Energy Engineering and International Research Center for Nuclei and Particles in the Cosmos, Beihang University, Beijing 100191, China
\and
School of Physics and State Key Laboratory of Nuclear Physics and Technology, Peking University, Beijing 100871, China
\and
Collaborative Innovation Center of Quantum Matter, Beijing 100871, China
\and
Center of High Energy Physics, Peking University, Beijing 100871, China}
\date{Received: date / Revised version: date}
\abstract{
To identify the nature of the newly observed charged resonance $Z_c(4200)^+$, we study its hadronic decays
$Z_c(4200)^+\to J/\psi\pi^+, Z_c(4200)^+\to\eta_c\rho^+$ and $Z_c(4200)^+\to D^+\bar D^{\ast 0}$ as
a charmonium-like tetraquark state. In the framework of the QCD sum rules, we calculate the three-point functions
and extract the coupling constants and decay widths for these interaction vertices. Including all these channels,
the full decay width of the $Z_c(4200)^+$ state is consistent with the experimental value reported by the Belle
Collaboration, supporting the tetraquark interpretation of this state.
 % end of PACS codes
} %end of abstract
\maketitle

%================================================================================
%================================================================================
%================================================================================
\section{Introduction}\label{sec:intro}
%================================================================================
%================================================================================

Recently, a new charged charmoniumlike resonance $Z_c(4200)^+$ was observed by the Belle Collaboration~\cite{2014-Chilikin-p112009-112009}.
It was observed in the $Z_c(4200)^+\to J/\psi\pi^+$ process with the mass and decay width $M=4196^{+31+17}_{-29-13}$ MeV
and $\Gamma=370^{+70+70}_{-70-132}$ MeV, with a significance of $6.2\sigma$. Its preferred assignment of the quantum
numbers is $J^P=1^+$. The G-parity of $Z_c(4200)^+$ is positive. Thus, the quantum numbers of its neutral partner is
$I^GJ^{PC}=1^+1^{+-}$.

The family of the charged charmoniumlike states have become more abundant after the discovery of $Z_c(4200)^+$~\cite{2014-Chilikin-p112009-112009}
and $Z_c(4050)$~\cite{2014-Wang-p-}.
Before this, the first member $Z(4430)^+$ was observed in the $\psi(2S)\pi^+$ invariant mass spectrum in the
process $\bar B^0\to \psi(2S)\pi^+K^-$ by the Belle Collaboration~\cite{2008-Choi-p142001-142001} and confirmed
recently by the LHCb Collaboration~\cite{2014-Aaij-p222002-222002}. Later, Belle also reported a broad doubly peaked
structure in the $\pi^+\chi_{c1}$ invariant mass distribution, which are called $Z(4050)^+$ and $Z(4250)^+$~\cite{2008-Mizuk-p72004-72004}.
Several other similar charged states were observed in last two years. In 2013, the BESIII Collaboration reported
$Z_c(3900)^+$ in $J/\psi\pi^+$ final states in the process $Y(4260)\to J/\psi\pi^+\pi^-$~\cite{2013-Ablikim-p252001-252001}.
$Z_c(3900)^+$ was also observed by Belle~\cite{2013-Liu-p252002-252002} and confirmed in CLEO data~\cite{2013-Xiao-p366-370}.
The BESIII Collaboration also observed $Z_c(4025)^{\pm}$ in the $\pi^{\mp}$ recoil mass spectrum in the
$e^+e^-\to (D^{\ast}\bar D^{\ast})^{\pm}\pi^{\mp}$ process~\cite{2014-Ablikim-p132001-132001} and $Z_c(4020)^{\pm}$
in the $h_c\pi^{\pm}$ mass spectrum in the process $e^+e^-\to h_c\pi^+\pi^-$~\cite{2013-Ablikim-p242001-242001}.
Moreover, the Belle Collaboration also observed two charged bottomoniumlike states $Z_b(10610)$ and $Z_b(10650)$
 in the $\pi^{\pm}\Upsilon(nS)$ and $h_b\pi^{\pm}$ mass spectra in the $\Upsilon(5S)$ decay~\cite{2011-Adachi-p-}.

These newly observed charged states have the exotic flavor contents $c\bar cu\bar d$ for $Z_c$ states and $b\bar bu\bar d$
for $Z_b$ states. It is natural to understand them as different manifestations of four-quark states: hadron molecules~\cite{Close:2003sg,Braaten:2003he,Tornqvist:2004qy,Swanson:2006st,Fleming:2007rp,Braaten:2007dw,Guo:2013sya}, tetraquark states~\cite{Maiani:2004vq,Ali:2011ug,Ebert:2008kb}, or many other configurations~\cite{Dubynskiy:2008mq,Danilkin:2011sh,Matheus:2009vq}.
For example, $Z(4430)^+$ was described as a $D^{\ast}\bar D_1$ molecular state in Refs.~\cite{2008-Liu-p34003-34003,2008-Lee-p28-32,2007-Meng-p-,2007-Ding-p-} and a tetraquark state in Refs.~\cite{2009-Bracco-p240-244,2008-Maiani-p73004-73004,2014-Maiani-p114010-114010}; the
$Z_c(3900)^+$ was speculated to be a molecular state in Refs.~\cite{2013-Wang-p132003-132003,2014-Aceti-p16003-16003,2014-Zhao-p94026-94026}; the
$Z_c(4025)^+$ was interpreted as a $D^*\bar D^*$ molecular state in Refs.~\cite{2013-He-p2635-2635}; the $Z_b(10610)$ and $Z_b(10650)$ were studied as $\bar BB^{\ast}$ and $\bar B^{\ast}B^{\ast}$ molecular states in Refs.~\cite{2009-Liu-p411-428}. One can consult Refs.~\cite{2014-Olive-p90001-90001,Brambilla:2010cs,Nielsen:2009uh,Esposito:2014rxa,Olsen:2014qna,Liu:2013waa} and references therein for recent reviews of these charged resonances.

Being composed of a diquark and antidiquark pair, a hidden-charm tetraquark state can decay very easily into a pair of open-charm
$D$ mesons or one charmonium state plus a light meson through quark rearrangement, implying that tetraquark states
should be very broad resonances while the experimental $XYZ$ states are usually quite narrow, such as
$Z_c(3900)^+$~\cite{2013-Ablikim-p252001-252001,2013-Liu-p252002-252002,2013-Xiao-p366-370}
and $Z_c(4025)^+$~\cite{2014-Ablikim-p132001-132001,2013-Ablikim-p242001-242001}.
However, the experimental width value of the $Z_c(4200)^+$~\cite{2014-Chilikin-p112009-112009}
is broad enough to be a good tetraquark candidate. In Ref.~\cite{2014-Zhao-p-}, $Z_c(4200)^+$ was studied as a tetraquark state
by considering the color-magnetic interaction. In Ref.~\cite{2014-Prelovsek-p-}, the authors tried to search for $Z_c^+$
exotic states in lattice QCD. However, they found no convincing signal for $Z_c^+$ state below $4.2$ GeV.

The hidden-charm tetraquark states with $J^{PC}=1^{+-}$ has been studied using the method of QCD sum rule in Refs.~\cite{Matheus:2009vq,Qiao:2013raa,2013-Cui-p-,2013-Dias-p16004-16004,Narison:2010pd,Zhang:2013aoa,Wang:2014gwa}.
We have also done similar QCD sum rule studies in Refs.~\cite{2011-Chen-p34010-34010,2012-Chen-p1003-1003}, in which the extracted mass was found to be consistent
with the experimental value of the $Z_c(4200)^+$ mass. In this work, we will study the hadronic decays of
the $Z_c(4200)^+$ as a tetraquark state in QCD sum rules. The three-point functions for the
$Z_cJ/\psi\pi, Z_c\eta_c\rho$ and $Z_cDD^{\ast}$ vertices will be studied to calculate the corresponding coupling constants
needed to extract the decay widths.

This paper is organized as follows. In Sec.~\ref{Sec:QSR}, we study the three-point functions for the $Z_cJ/\psi\pi$, $Z_c\eta_c\rho$
and $Z_cDD^{\ast}$ vertices. We will calculate the OPE series up to dimension five condensates. Then we compute the coupling
constants and the decay widths for these channels. Finally, we give a short summary and discuss the possibility of searching
for such exotic resonances decaying into $\eta_c$ charmonium.

%================================================================================
%================================================================================
\section{QCD sum rules and three-point correlation function}\label{Sec:QSR}
%================================================================================
%================================================================================
In the past several decades, QCD sum rule has proven to be a very powerful non-perturbative approach to
study hadron properties such as masses, magnetic moments and coupling constants, associated with
the low-lying baryons and mesons~\cite{1979-Shifman-p385-447,1985-Reinders-p1-1,2000-Colangelo-p1495-1576,Ioffe-1983ju,Eidemuller-2005jm}.
Recently, this method was used to yield predictions on the spectroscopy of the new hadron $XYZ$
states~\cite{Matheus:2009vq,Nielsen:2009uh,Qiao:2013raa,2013-Cui-p-,2013-Dias-p16004-16004,Narison:2010pd,Zhang:2013aoa,Wang:2014gwa,2011-Chen-p34010-34010,2012-Chen-p1003-1003,2014-Chen-p13-40}.

To calculate the decay width of the $Z_c(4200)$ meson into two hadrons, one needs first to study the three-body coupling vertices
$Z_c(4200)AB$, where $A$, $B$ denote the decay products. In QCD sum rules, we consider the three-point correlation function
\begin{eqnarray}
&& \Pi_{\mu\nu}(p, p^{\prime}, q)
\label{three_point_function}
\\ \nonumber && ~~~ = \int
d^4xd^4y \,e^{ip^{\prime} \cdot x}e^{iq \cdot y}\,\langle0|T[J^A(x)J^B(y)J_{\nu}^{Z_c\dag}(0)]|0\rangle,
\end{eqnarray}
where $J_{\nu}^{Z_c}$ is the interpolating current for the $Z_c(4200)^+$ meson while $J^A$ and $J^B$ are the currents
for the final states $A$ and $B$, respectively. In this paper, we consider the $Z_c(4200)^+$ meson as a charmonium-like tetraquark state.
The corresponding tetraquark current is given by
\begin{eqnarray}
%J_{\mu}^{\psi}&=&\bar c_a\gamma_{\mu}c_a,
%\\
%J^{\pi}&=&\bar d_a\gamma_5 u_a,
%\\
\label{eq:zc}
J_{\nu}^{Z_c}&=&u^T_{a}C\gamma_5c_{b}(\bar{d}_{a}\gamma_{\nu}C\bar{c}^T_{b}+\bar{d}_{b}\gamma_{\nu}C\bar{c}^T_{a})
\\ \nonumber &&- u^T_{a}C\gamma_{\nu}c_{b}(\bar{d}_{a}\gamma_{5}C\bar{c}^T_{b}+\bar{d}_{b}\gamma_{5}C\bar{c}^T_{a}),
\end{eqnarray}
in which the subscripts $a, b$ are color indices, and $u, d$ and $c$ represent up, down and charm quarks, respectively.
$C$ is the charge-conjugation matrix. We have studied this charmoniun-like tetraquark scenario and the extracted mass is
around $4.16$ GeV~\cite{2011-Chen-p34010-34010} consistent with the observed mass
of the $Z_c(4200)^+$ meson~\cite{2014-Chilikin-p112009-112009}. This current can couple to the $Z_c(4200)^+$ meson via
\begin{eqnarray}
\langle 0|J_{\nu}^{Z}|Z_c(p)\rangle&=&f_{Z}\epsilon_{\nu}(p), \label{Zcouple}
\end{eqnarray}
in which $\epsilon_{\nu}(p)$ is a polarization vector and $f_Z$ is the coupling constant of the current to the physical state.

The $Z_c(4200)$ meson can decay into several different channels such as hidden-charm decay modes $J/\psi\pi^+, \eta_c\rho^+$
and open-charm decay modes $D^+\bar D^{\ast 0}, \bar D^0D^{\ast +}$.
Such decay properties are similar to those for the
charmonium-like state $Z_c(3900)$. Assuming $Z_c(3900)$ to be a tetraquark state with the same quantum numbers as the $Z_c(4200)$,
the hadronic decay modes of $Z_c(3900)$ to $J/\psi\pi^+, \eta_c\rho^+, D^+\bar D^{\ast 0}$ and $\bar D^0D^{\ast +}$ were studied
in Ref.~\cite{2013-Dias-p16004-16004}. Building upon these methods, we will study the same decay channels for the $Z_c(4200)$
tetraquark state to estimate its decay width.

%================================================================================
\subsection{Decay mode $Z_c^+(4200)\to J/\psi\pi^+$}\label{Sec:psipi}
%================================================================================

In this subsection, we study the hidden-charm decay $Z_c^+(4200)\to J/\psi\pi^+$, in which mode the $Z_c^+(4200)$
meson was observed. To calculate the three-point function in Eq.~\eqref{three_point_function} for the vertex
$Z_c^+(4200)J/\psi\pi^+$, we need the interpolating currents for $J/\psi$ and $\pi$ mesons
\begin{eqnarray}
J_{\mu}^{\psi}&=&\bar c_a\gamma_{\mu}c_a,
\label{eq:psi}
\\
J^{\pi}&=&\bar d_a\gamma_5 u_a,
\label{eq:pi}
%\\
%J_{\nu}^{Z_c}&=&u^T_{a}C\gamma_5c_{b}(\bar{d}_{a}\gamma_{\nu}C\bar{c}^T_{b}+\bar{d}_{b}\gamma_{\nu}C\bar{c}^T_{a})
%-u^T_{a}C\gamma_{\nu}c_{b}(\bar{d}_{a}\gamma_{5}C\bar{c}^T_{b}+\bar{d}_{b}\gamma_{5}C\bar{c}^T_{a}),
\end{eqnarray}
which can couple to the $J/\psi$ and $\pi$ mesons respectively via the following relations
\begin{eqnarray}
\langle0|J_{\mu}^{\psi}|J/\psi(p^{\prime})\rangle&=&f_{\psi}\epsilon_{\mu}(p^{\prime}),
\\
\langle0|J^{\pi}|\pi(q)\rangle&=&f_{\pi}^{\prime}=\frac{2i\qq}{f_{\pi}},
%\\
%\langle Z_c(p)|J_{\nu}^{\dag Z}|0\rangle&=&f_{Z}\epsilon_{\nu}^{\ast}(p),
\end{eqnarray}
where $f_{\psi}$ and $f_{\pi}^{\prime}$ are coupling constants.
Using these relations and Eq.~\eqref{Zcouple}, we can write down the three-point function in the phenomenological side
\begin{eqnarray}
&& \Pi_{\mu\nu}^{\psi\pi}(p, p^{\prime}, q)
\label{PHpi}
\\ \nonumber &=&\int
d^4xd^4y \,e^{ip^{\prime} \cdot x}e^{iq \cdot y}\,\langle0|T[J_{\mu}^{\psi}(x)J^{\pi}(y)J_{\nu}^{Z\dag}(0)]|0\rangle
\non &=& g_{Z\psi\pi}g_{\mu^{\prime}\nu^{\prime}}\left(g_{\mu\mu^{\prime}}-\frac{p^{\prime}_{\mu}p^{\prime}_{\mu^{\prime}}}{m_{\psi}^2}\right)
\left(g_{\nu\nu^{\prime}}-\frac{p_{\nu}p_{\nu^{\prime}}}{m_{Z}^2}\right) \times
\\ \nonumber && \frac{f_{\psi}(-if_{\pi}^{\prime})f_{Z}}{(p^2-m^2_{Z}+i\epsilon)(p^{\prime 2}-m^2_{\psi}+i\epsilon)(q^2-m^2_{\pi}+i\epsilon)}+...
\non &=&\frac{g_{Z\psi\pi}(q^2)f_{\psi}(-if_{\pi}^{\prime})f_{Z}}{(p^2-m^2_{Z}+i\epsilon)(p^{\prime 2}-m^2_{\psi}+i\epsilon)(q^2-m^2_{\pi}+i\epsilon)}\times
\\&& \nonumber
\left(g_{\mu\nu}-\frac{q_{\mu}p^{\prime}_{\nu}+q_{\mu}q_{\nu}}{m_Z^2}-\frac{p^{\prime}_{\mu}p^{\prime}_{\nu}}{m_{\psi}^2}+
\frac{p^{\prime}\cdot q(p^{\prime}_{\mu}p^{\prime}_{\nu}+q_{\nu}p^{\prime}_{\mu})}{m_Z^2m_{\psi}^2}\frac{}{}\right)
\\ \nonumber && +...,
\end{eqnarray}
in which ``$...$'' represents the contributions of all higher excited states. We have used the relation $p=p^{\prime}+q$ in the last step.
The coupling constant (form factor) $g_{Z\psi\pi}(q^2)$ is defined via
\begin{eqnarray}
\mathcal{L}=g_{Z\psi\pi}g^{\mu\nu}Z_{c\nu}^{+}\pi^-\psi_{\mu}+h.c..
\end{eqnarray}
%In the following analysis, we extract the coupling constant $g_{Z\psi\pi}$ instead of the form factor.
In Eq.~\eqref{PHpi}, the three-point function $\Pi_{\mu\nu}^{\psi\pi}(p, p^{\prime}, q)$ is divergent at $q^2=0$ when we take the limit $m_{\pi}=0$.
To simplify the calculation in OPE side, we establish a sum rule at the massless pion-pole, which was first suggested in
Ref.~\cite{1985-Reinders-p1-1} for the pion nucleon coupling constant.

%\cofeBm{1}{1.0}{0}{11.5cm}{2cm}
%\cofeCm{1}{1.0}{0}{5.5cm}{15cm}

At the quark-gluon level, the three-point function in Eq.~\eqref{three_point_function} can be evaluated via the operator product expansion (OPE) method.
We insert the three currents, Eqs.~(\ref{eq:zc}), (\ref{eq:psi}) and (\ref{eq:pi}), into the three-point function, Eq.~(\ref{PHpi}), and do the Wick contraction:
\begin{eqnarray} \label{contract}
&&\langle0|T[J_{\mu}^{\psi}(x)J^{\pi}(y)J_{\nu}^{Z\dag}(0)]|0\rangle
\\ \nonumber &=& -{\bf Tr} [ S^c_{ab'}(x)\gamma_{5} S^{'u}_{ba'}(y)\gamma_5 S^{'d}_{a'b}(-y)\gamma_{\nu}S^c_{b'a}(-x)\gamma_{\mu}]
\non && -{\bf Tr} [ S^c_{ab'}(x)\gamma_{5} S^{'u}_{ba'}(y)\gamma_5 S^{'d}_{b'b}(-y)\gamma_{\nu}S^c_{a'a}(-x)\gamma_{\mu}]
\non && -{\bf Tr} [ S^c_{ab'}(x)\gamma_{\nu} S^{'u}_{ba'}(y)\gamma_5 S^{'d}_{a'b}(-y)\gamma_{5}S^c_{b'a}(-x)\gamma_{\mu}]
\\ \nonumber && -{\bf Tr} [ S^c_{ab'}(x)\gamma_{\nu} S^{'u}_{ba'}(y)\gamma_5 S^{'d}_{b'b}(-y)\gamma_{5}S^c_{a'a}(-x)\gamma_{\mu}],
\end{eqnarray}
where the subscripts $a, b, a^{\prime}$ and
$b^{\prime}$ are color indices, and the superscripts $u, d$ and $c$
denote the quark propagators for up, down and charm quark, respectively. Throughout our evaluation,
we use the coordinate-space expression for the light quark propagator and
momentum-space expression for the heavy quark propagator~\cite{1985-Reinders-p1-1,1993-Yang-p3001-3012}:
\begin{eqnarray} \nonumber
iS_{ab}^q(x)&=& \frac{i\delta_{ab}}{2\pi^2x^4}\hat{x}
+\frac{i}{32\pi^2}\frac{\lambda^n_{ab}}{2}gG_{\mu\nu}^n\frac{1}{x^2}(\sigma^{\mu\nu}\hat{x}+\hat{x}\sigma^{\mu\nu})
\\ &&-\frac{\delta_{ab}}{12}\langle\bar{q}q\rangle + \frac{\delta_{ab}x^2}{192}\langle g_s \bar{q}\sigma Gq\rangle, \label{coordinate_propagator}
\\ \nonumber
iS_{ab}^c(p) &=&
\frac{i\delta_{ab}(p\!\!\!\slash+m_c)}{p^2-m_c^2}+\frac{i}{4}g\frac{\lambda^n_{ab}}{2}G^n_{\mu\nu}\frac{1}{(p^2-m_c^2)^2} \times
\\ && \{\sigma_{\mu\nu}(p\!\!\!\slash+m_c)+(p\!\!\!\slash+m_c)\sigma_{\mu\nu}\}, \label{momentum_propagator}
\end{eqnarray}
where $m_c$ is the mass of the charm quark. We neglect the chirally-suppressed contributions from the current quark masses ($m_q=0$ in the chiral limit) because they are numerically insignificant. In Eq.~\eqref{contract}, the light quark propagator is defined as $iS_{ab}^{'q}(x)=C(iS_{ab}^q)^TC$ in which $T$ represents only the transpose operation to the Dirac indices. As indicated above, we will pick out
the $1/q^2$ terms in the OPE series and work at the limit $q^2\to 0$. We note that this is the assumption used in Ref.~\cite{1985-Reinders-p1-1}, and then we can establish a sum rule by comparing with the three-point
function expression Eq.~\eqref{PHpi} at the hadron level.

%\cofeBm{0.5}{0.8}{0}{7.0cm}{9cm}
% Our calculation shows that both the perturbative
% and the gluon condensate terms contain twice divergences in the three-point function. In general, one needs to perform twice Borel transforms to
% eliminate these divergences. It is very complicated because that will give two Borel parameters. However, this problem will not appear when we work
% at the pion pole as mentioned above. The perturbative term and the gluon condensate give no contributions to the $1/q^2$ proportional terms. We just
% need to consider the quark condensate and quark-gluon mixed condensate.

\begin{figure*}
\begin{center}
\resizebox{0.50\textwidth}{!}{%
  \includegraphics{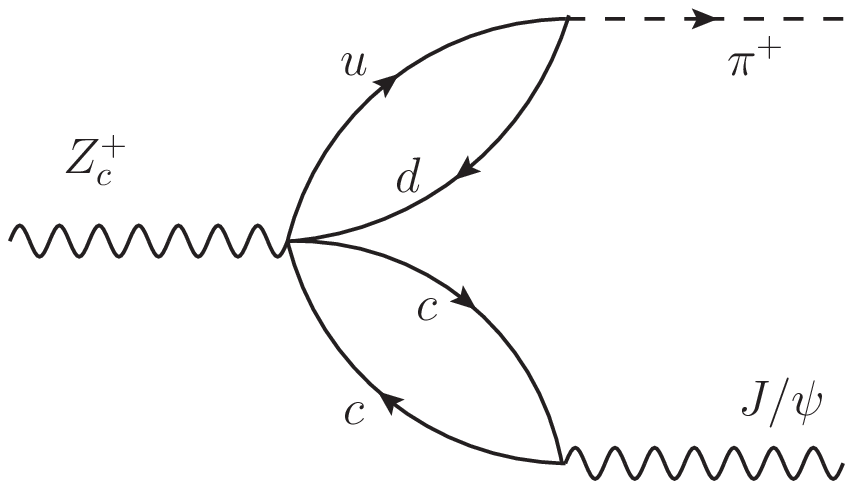}
  \includegraphics{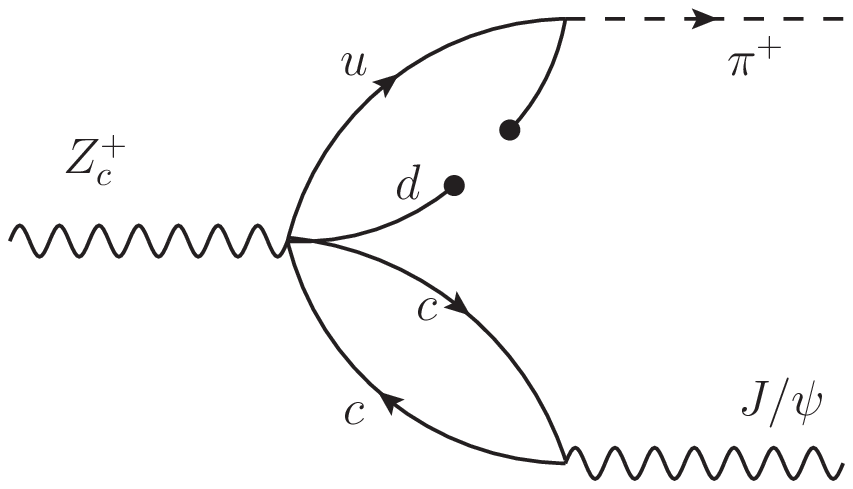}}
  \\[1cm]
\resizebox{0.75\textwidth}{!}{
  \includegraphics{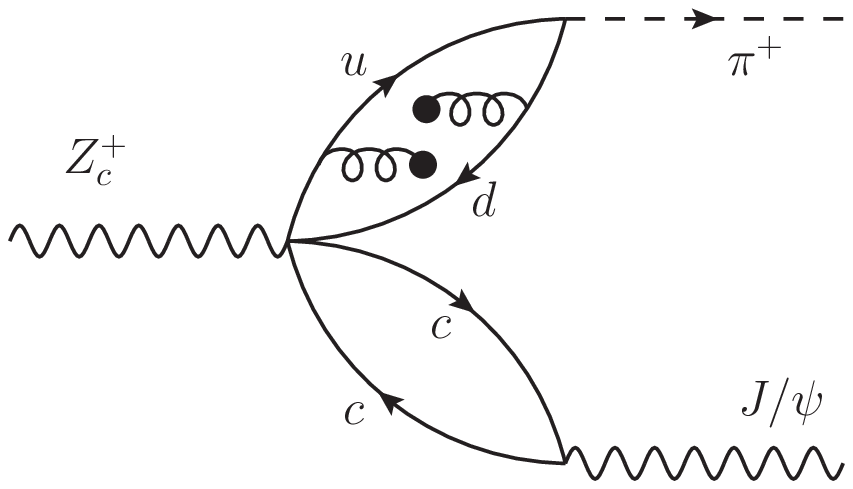}
  \includegraphics{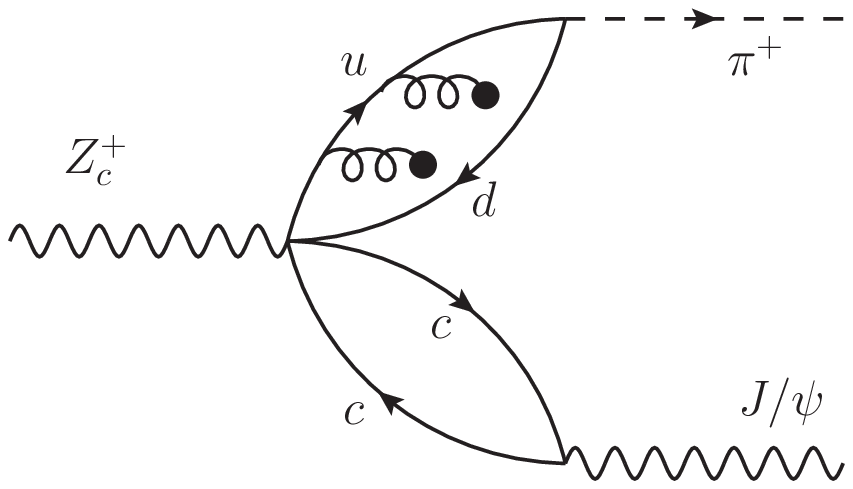}
  \includegraphics{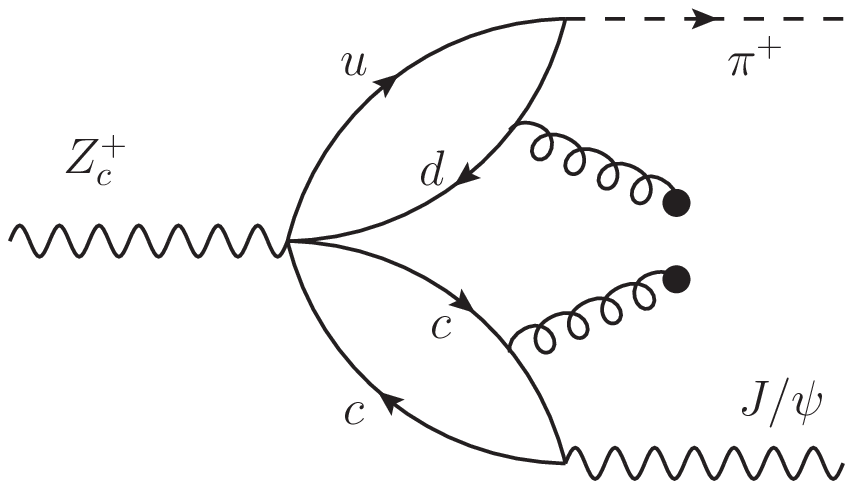}}
  \\[1cm]
\resizebox{0.75\textwidth}{!}{
  \includegraphics{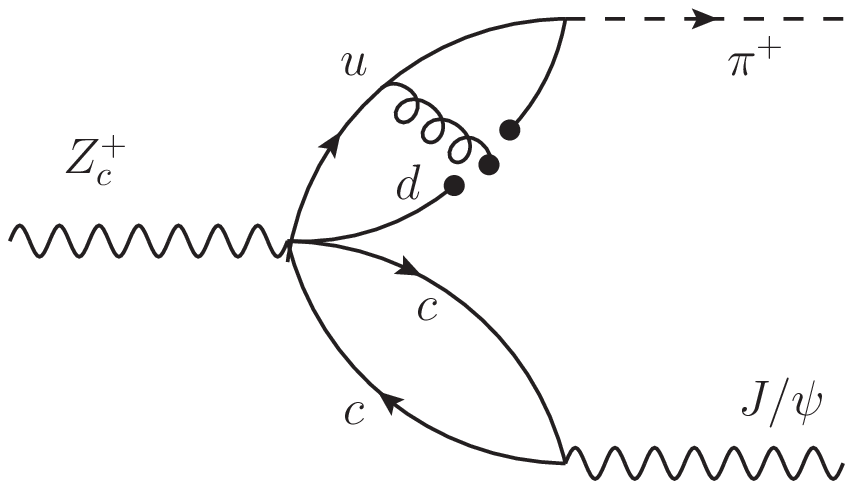}
  \includegraphics{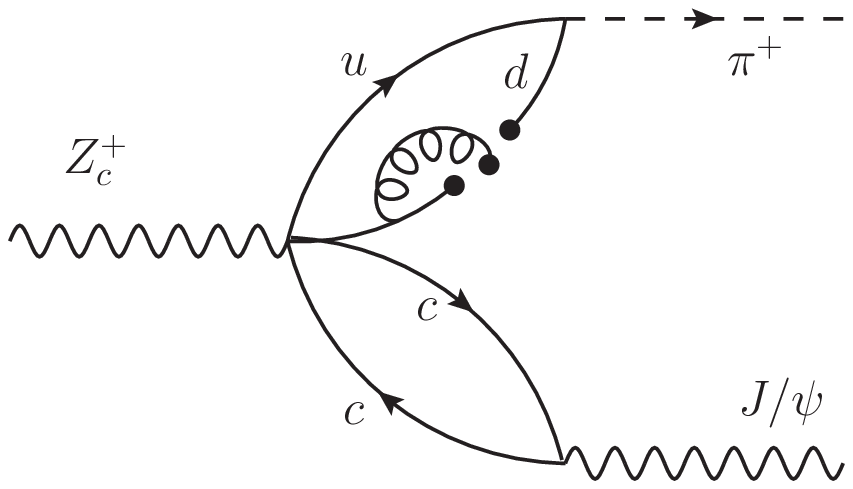}
  \includegraphics{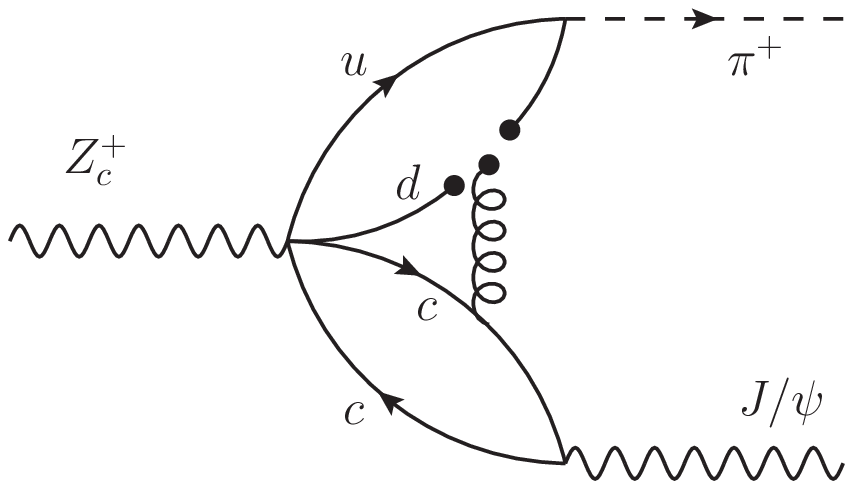}}       % Give the correct figure height in cm
\caption{Feynman diagrams for the leading order contributions of the three-point function. Wave lines depict vector and axial-vector mesons,
dashed lines pion, solid lines quarks and curly lines gluons. Graphs related by symmetry are not shown.}
\label{Graphs}       % Give a unique label
\end{center}
\end{figure*}

On the phenomenological side in Eq.~\eqref{PHpi}, there are five different tensor structures $g_{\mu\nu}, q_{\mu}q_{\nu}, q_{\mu}p^{\prime}_{\nu}, q_{\nu}p^{\prime}_{\mu}$ and $p^{\prime}_{\mu}p^{\prime}_{\nu}$. On the QCD side, we evaluate the three-point function and spectral
density up to the dimension five terms. In addition to the perturbative term, we calculate the quark condensate, the gluon condensate and the quark-gluon mixed condensate for the power corrections. The Feynman diagrams for these terms are shown in Fig.~\ref{Graphs}. In our result for the OPE series, the $g_{\mu\nu}$ structure contributes to
all expansion terms including the perturbative part, quark condensate, gluon condensate and quark-gluon mixed condensate. Other tensor structures
contribute just some of these terms in the OPE at leading order. For example, the $q_{\mu}q_{\nu}$ and $q_{\nu}p^{\prime}_{\mu}$ structures appear only in the gluon condensate while $p^{\prime}_{\mu}p^{\prime}_{\nu}$ appears in the perturbative term and the gluon condensate. The structure $q_{\mu}p^{\prime}_{\nu}$ gives no contributions to the perturbative term.

To obtain the greatest number of terms in the OPE series, we therefore study the
$g_{\mu\nu}$ structure in the following analysis. Finally, we obtain the spectral density proportional to $1/q^2$ in the $g_{\mu\nu}$ structure
\begin{eqnarray}
\rho(s)=-\frac{\GGb (s+2m_c^2)}{192\pi^4}\sqrt{1-\frac{4m_c^2}{s}}, \label{spectral_density}
\end{eqnarray}
where we find that only the gluon condensate gives contributions to the spectral density at order  $1/q^2$.
There are three kinds Feynman diagrams for the gluon condensate in Fig. \ref{Graphs}, in which the third one is color-connected and the former two are color-disconnected
\cite{2013-Dias-p16004-16004}. In our caculation, the spectral density in Eq. \eqref{spectral_density} comes from only the first color-disconnected diagram in Fig. \ref{Graphs}.
The second color-disconnected diagram and the color-connected diagrams give no contribution to $\rho(s)$.

To establish a sum rule for the coupling constant $g_{Z\psi\pi}$, we assume $p^2=p^{\prime 2}=P^2$ in Eq.~\eqref{PHpi} and then perform the
Borel transform ($P^2\to M_B^2$) to suppress the higher state contributions. For the $g_{\mu\nu}$ structure, we arrive at the sum rule
%\begin{eqnarray}
%\frac{g_{Z\psi\pi}(q^2)f_{\psi}(-if_{\pi}^{\prime})f_{Z}}{m^2_{Z}-m^2_{\psi}}\left(e^{-m^2_{\psi}/M_B^2}-e^{-m^2_{Z}/M_B^2}\right)=\int_{4m_c^2}^{\infty}\rho(s)e^{-s/M_B^2}ds
%\end{eqnarray}
\begin{eqnarray}
\label{g_psipi}
&& g_{Z\psi\pi}(s_0, M_B^2)|_{Q^2\to 0}
\non && =\frac{1}{f_{\psi}(-if^{\prime}_{\pi})f_{Z}}\frac{m^2_{Z}-m^2_{\psi}}{e^{-m^2_{\psi}/M_B^2}-e^{-m^2_{Z}/M_B^2}}\int_{4m_c^2}^{s_0}\rho(s)e^{-s/M_B^2}ds,
\end{eqnarray}
in which $Q^2=-q^2$ and $s_0$ is the continuum threshold parameter for the $Z_c(4200)$ meson.

To perform the QCD sum rule numerical analysis, we use the following values of quark masses and various condensates~\cite{2014-Olive-p90001-90001,1985-Reinders-p1-1,2012-Narison-p259-263,1997-Narison-p238-243}:
\begin{eqnarray}
\nonumber
&&m_c(\mu=m_c)=\overline m_c=(1.275\pm 0.025)~\mbox{GeV}   \, ,
\non &&\qq=-(0.23\pm0.03)^3\text{ GeV}^3 \, ,
\\   &&\qGqb=-M_0^2\qq\, , \label{parameters_QCD}
\non &&M_0^2=(0.8\pm0.2)\text{ GeV}^2 \, ,
\non &&\GGb=(0.48\pm0.14)\text{GeV}^4\, ,
\end{eqnarray}
in which the charm quark mass is the running mass in the $\overline{\rm MS}$ scheme.
Note that there is a minus sign implicitly included in the definition of the coupling constant $g_s$ in this work.
We use the following values of the hadron
parameters~\cite{1985-Reinders-p1-1,2014-Olive-p90001-90001,2014-Chilikin-p112009-112009,2011-Chen-p34010-34010}
\begin{eqnarray}
\nonumber
m_{\psi}&=&(3.097\pm0.011)~\mbox{GeV}, ~f_{\psi}=(1.288\pm0.037)~\mbox{GeV}^2,
\\ m_{\pi}&=&139.6~\mbox{MeV}, ~f_{\pi}=133~\mbox{MeV}, \label{parameters_psipi}
\non m_{Z_c}&=&(4.196^{+0.048}_{-0.042})~\mbox{GeV}, ~f_{Z_c}=(6.9\pm0.4)\times 10^{-3}~\mbox{GeV}^5,
\end{eqnarray}
in which the coupling parameter $f_{Z_c}$ is determined by the mass sum rules in Ref.~\cite{2011-Chen-p34010-34010}.

\begin{figure}
\begin{center}
\resizebox{0.40\textwidth}{!}{%
  \includegraphics{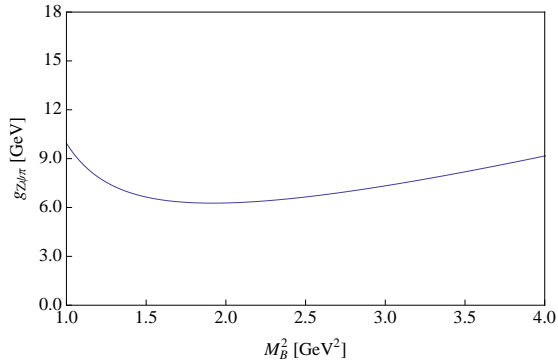}}       % Give the correct figure height in cm
\caption{Coupling constant $g_{Z\psi\pi}$ dependence on the Borel parameter $M_B^2$.}
\label{fig_gpsipi}       % Give a unique label
\end{center}
\end{figure}

In Eq.~\eqref{g_psipi}, the coupling constant $g_{Z\psi\pi}(q^2)$ depends on the continuum threshold value $s_0$
and the Borel mass $M_B$ in the limit $q^2\to 0$. We use the continuum threshold value $s_0=21$ GeV$^2$,
which is the same value as used in the mass sum rule in Ref.~\cite{2011-Chen-p34010-34010}. Using this value
of $s_0$ and the parameters given in Eqs.~\eqref{parameters_QCD} and \eqref{parameters_psipi}, we show the
variation of the coupling constant $g_{Z\psi\pi}$ with the Borel parameter $M_B^2$ in Fig.~\ref{fig_gpsipi}.
We find that the sum rule gives a minimum (stable) value of the coupling constant $g_{Z\psi\pi}=(6.27\pm1.93)$ GeV
with $M_B^2\sim1.9$ GeV$^2$. The errors come from the uncertainties of the charm quark mass, the gluon
condensate, hadron masses and hadron couplings.
To calculate the decay width of a process $A\to BC$, we use the following relation given by
Refs.~\cite{2013-Maiani-p111102-111102,2013-Dias-p16004-16004}:
\begin{eqnarray}
&& \Gamma(A\to BC) \label{decaywidth}
\non &&=\frac{p^*(m_A, m_B, m_C)}{8\pi m_A^2}\times\frac{g^2_{ABC}}{3}
\left(3+\frac{p^*(m_A, m_B, m_C)^2}{m_B^2}\right),
\end{eqnarray}
where $g_{ABC}$ is the coupling of the three-point vertex $ABC$ and $p^*(m_A, m_B, m_C)$ is defined as
\begin{eqnarray}
p^*(a, b, c)=\frac{\sqrt{a^4+b^4+c^4-2a^2b^2-2b^2c^2-2c^2a^2}}{2a}.
\end{eqnarray}

For the decay mode $Z_c^+(4200)\to J/\psi\pi^+$, we then can calculate its decay width using Eq.~\eqref{decaywidth}
\begin{eqnarray}
\Gamma(Z_c^+(4200)\to J/\psi\pi^+)=(87.3\pm47.1) ~\mbox{MeV}, \label{width_psipi}
\end{eqnarray}
in which the dominate error source is the uncertainty of the gluon condensate.

%================================================================================
\subsection{Decay mode $Z_c^+(4200)\to \eta_c\rho^+$}\label{Sec:etarho}
%================================================================================
We study the decay mode $Z_c^+(4200)\to \eta_c\rho^+$ in this subsection. To calculate the corresponding
three-point function, we consider the following interpolating currents for $\eta_c$ and $\rho$ mesons
\begin{eqnarray}
J^{\eta_c}&=&\bar c_a\gamma_{5}c_a,
\\
J_{\mu}^{\rho}&=&\bar d_a\gamma_{\mu} u_a,
\end{eqnarray}
with the current-meson coupling relation
\begin{eqnarray}
\langle0|J^{\eta_c}|\eta_c(p^{\prime})\rangle&=&f^{\prime}_{\eta_c}=-i\frac{f_{\eta_c}m_{\eta_c}^2}{2m_c},
\\
\langle0|J_{\mu}^{\rho}|\rho(q)\rangle&=&f_{\rho}\epsilon_{\mu}(q),
\end{eqnarray}
where $f_{\eta_c}$ and $f_{\rho}$ are coupling constants for $\eta_c$ and $\rho$, respectively. Then we can
obtain the three-point function at the hadron level
\begin{eqnarray}
\label{PHpi_etarho}
&& \Pi_{\mu\nu}^{\eta_c\rho}(p, p^{\prime}, q)
\non &=&\int
d^4xd^4y \,e^{ip^{\prime} \cdot x}e^{iq \cdot y}\,\langle0|T[J^{\eta_c}(x)J^{\rho}_{\mu}(y)J_{\nu}^{Z\dag}(0)]|0\rangle
\non &=&\frac{g_{Z\eta_c\rho}(q^2)f_{\rho}(-if^{\prime}_{\eta_c})f_{Z}}{(p^2-m^2_{Z}+i\epsilon)(p^{\prime 2}-m^2_{\eta_c}+i\epsilon)(q^2-m^2_{\rho}+i\epsilon)}\times
\\&&\nonumber
\left(g_{\mu\nu}-\frac{q_{\nu}p^{\prime}_{\mu}+p^{\prime}_{\mu}p^{\prime}_{\nu}}{m_Z^2}-\frac{q_{\mu}q_{\nu}}{m_{\rho}^2}+
\frac{p^{\prime}\cdot q(q_{\mu}q_{\nu}+q_{\mu}p^{\prime}_{\nu})}{m_Z^2m_{\rho}^2}\right)
\non &&+...,
\end{eqnarray}
where the coupling constant $g_{Z\eta_c\rho}(q^2)$ is defined via
\begin{eqnarray}
 \mathcal{L}=g_{Z\eta_c\rho}g^{\mu\nu}Z_{c\nu}^{+}\eta_c\rho_{\mu}^-+h.c..
\end{eqnarray}

At the quark-gluon level, we calculate the three-point correlation function $\Pi_{\mu\nu}^{\eta_c\rho}(p, p^{\prime}, q)$
by considering diagrams similar to Fig.~\ref{Graphs}.
As outlined in Ref.~\cite{2013-Dias-p16004-16004}, the coupling constant varies slowly with the Euclidean momentum $Q^2=-q^2$,
and hence for sufficiently large $Q^2$ it is only necessary to extract the $1/Q^2$ term in
Eq.~\eqref{PHpi_etarho} for the three-point correlation function $\Pi_{\mu\nu}^{\eta_c\rho}(p, p^{\prime}, q)$.
%\tcomm{I think you are making the argument that for large $q^2$ that we can neglect $m_\rho$?  I think we need to discuss this section before I can give more comments. }
Thus, for an appropriate range of $Q^2$ we keep
the invariant function proportional to $1/Q^2$ in the $g_{\mu\nu}$ structure. In this assumption \cite{2013-Dias-p16004-16004}, we obtain the spectral density
on the OPE side
\begin{eqnarray}
\rho(s)=-\left(\frac{m_c\qGqb}{6\pi^2}+\frac{\GGb s}{192\pi^4}\right)\sqrt{1-\frac{4m_c^2}{s}}, \label{spectral_density_etarho}
\end{eqnarray}
in which only the gluon condensate and quark-gluon mixed condensate give contributions to the spectral density in this order.
Similarly to the spectral density in Eq. \eqref{spectral_density},
the color-connected diagrams in Fig. \ref{Graphs} give no contributions to this spectral density in Eq.~\eqref{spectral_density_etarho}.
Then the sum rule for the coupling constant $g_{Z\eta_c\rho}$ can be established by assuming $p^2=p^{\prime 2}=P^2$
and performing the Borel transform ($P^2\to M_B^2$)
\begin{eqnarray}
\nonumber g_{Z\eta_c\rho}(s_0, M_B^2, Q^2)&=& \frac{1}{f_{\rho}(-if^{\prime}_{\eta_c})f_{Z}}\frac{m^2_{Z}-m^2_{\eta_c}}{e^{-m^2_{\eta_c}/M_B^2}-e^{-m^2_{Z}/M_B^2}}
\non && \times \left(\frac{Q^2+m_\rho^2}{Q^2}\right)\int_{4m_c^2}^{s_0}\rho(s)e^{-s/M_B^2}ds,
\\ \label{g_etarho}
\end{eqnarray}
in which $Q^2=-q^2$.

To perform the numerical analysis, we use the following parameters for the $\eta_c$ and $\rho$
mesons~\cite{2014-Olive-p90001-90001,2013-Dias-p16004-16004}
\begin{eqnarray}
\nonumber
m_{\eta_c}&=&(2.980\pm0.001)~\mbox{GeV}, ~f_{\eta_c}=0.35~\mbox{GeV},
\\ m_{\rho}&=&(775.26\pm0.25)~\mbox{MeV}, ~f_{\rho}=157~\mbox{MeV}. \label{parameters_etarho}
\end{eqnarray}

\begin{figure}
\begin{center}
\resizebox{0.40\textwidth}{!}{%
  \includegraphics{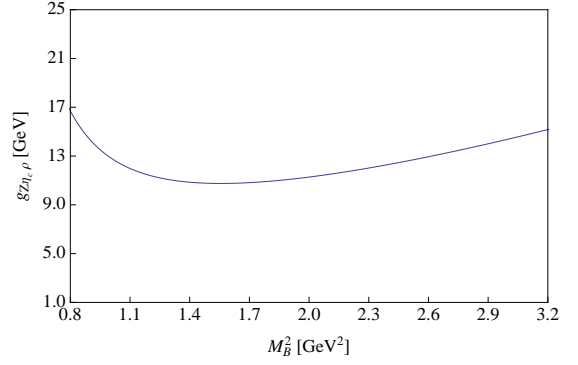}}       % Give the correct figure height in cm
\caption{Variation of the coupling constant $g_{Z\eta_c\rho}(Q^2)$ with the Borel parameter $M_B^2$ for $Q^2=8$ GeV$^2$.}
\label{fig_getarho}       % Give a unique label
\end{center}
\end{figure}

\begin{figure}
\begin{center}
\resizebox{0.40\textwidth}{!}{%
  \includegraphics{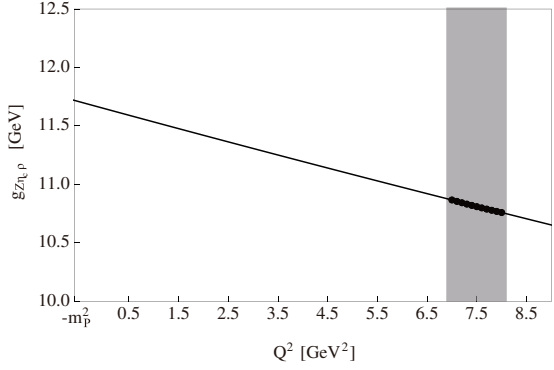}}       % Give the correct figure height in cm
\caption{Variation of the coupling constant $g_{Z\eta_c\rho}(Q^2)$ with $Q^2$. Data points shows the
QCD sum-rule result with $s_0=21$ GeV$^2$ and $M_B^2=1.6$ GeV$^2$. Solid line gives the fit of the QCD sum-rule result through Eq.~\eqref{model}
and the extrapolation of the coupling constant $g_{Z\eta_c\rho}(Q^2)$ to the physical pole $Q^2=-m_\rho^2$.}
\label{fig_getarho_Q}       % Give a unique label
\end{center}
\end{figure}

In Fig.~\ref{fig_getarho}, we show the variation of $g_{Z\eta_c\rho}(Q^2)$ with the Borel parameter for $Q^2=8$ GeV$^2$.
We choose $m_{\rho}^2\ll Q^2\sim m_{\eta_c}^2$ so that the $m_{\rho}$ can be ignored and the OPE is valid.
It shows that the minimum value of $g_{Z\eta_c\rho}$ appears around $M_B^2=1.6$ GeV$^2$. However, the
coupling constant is required at the pole $Q^2=-m_\rho^2$, where the QCD sum rule is not valid. To obtain the
coupling constant $g_{Z\eta_c\rho}(Q^2=-m_\rho^2)$, we need to extrapolate the coupling constant $g_{Z\eta_c\rho}(Q^2)$ from
the QCD sum-rule region to the physical pole $Q^2=-m_\rho^2$. Following Ref.~\cite{2013-Dias-p16004-16004}, we use the
exponential model to achieve this extrapolation
\begin{eqnarray}
g_{Z\eta_c\rho}(Q^2)=g_1e^{-g_2Q^2}, \label{model}
\end{eqnarray}
where $g_1$ and $g_2$ can be determined by fitting the QCD sum-rule result of $g_{Z\eta_c\rho}(Q^2)$, using Eq.~\eqref{model}.
In Fig.~\ref{fig_getarho_Q}, we show the QCD sum-rule result of the coupling constant $g_{Z\eta_c\rho}(Q^2)$ with $s_0=21$ GeV$^2$
and $M_B^2=1.6$ GeV$^2$ (dotted line). We fit this result by using the model Eq.~\eqref{model} with the parameters
$g_1=11.65$ GeV and $g_2=9.93\times10^{-3}$ GeV$^2$. Then we extrapolate the coupling constant $g_{Z\eta_c\rho}(Q^2)$
to the physical pole $Q^2=-m_\rho^2$ to obtain the coupling constant
\begin{eqnarray}
g_{Z\eta_c\rho}(Q^2=-m_\rho^2)=(11.72\pm2.10)~\mbox{GeV}, \label{gvalue_etarho}
\end{eqnarray}
in which the uncertainties of the charm quark mass, the QCD condensates, the hadron masses and the hadron couplings
are considered to give the error of this prediction.
Inserting this value into Eq.~\eqref{decaywidth}, we can calculate the decay width of the process
$Z_c^+(4200)\to \eta_c\rho^+$
\begin{eqnarray}
\Gamma(Z_c^+(4200)\to \eta_c\rho^+)=(334.4\pm119.8) ~\mbox{MeV}. \label{width_etarho}
\end{eqnarray}

%================================================================================
\subsection{Decay mode $Z_c^+(4200)\to D^+\bar{D}^{\ast 0}$}\label{Sec:DD}
%================================================================================
In this subsection, we study the open-charm decay mode $Z_c^+(4200)\to D^+\bar{D}^{\ast 0}$ with the
interpolating currents for the $D^+$ and $\bar{D}^{\ast 0}$ mesons
\begin{eqnarray}
J^{D^+}&=&\bar d_a\gamma_{5}c_a,
\\
J_{\mu}^{\bar{D}^{\ast0}}&=&\bar c_a\gamma_{\mu} u_a,
\end{eqnarray}
with the current-meson coupling relations
\begin{eqnarray}
\langle0|J^{D^+}|D^+(p^{\prime})\rangle&=&f_{D},
\\
\langle0|J_{\mu}^{\bar{D}^{\ast0}}|D^{\ast0}(q)\rangle&=&f_{D^{\ast}}\epsilon_{\mu}(q),
\end{eqnarray}
where $f_{D}$ and $f_{D^{\ast}}$ are the current-meson coupling constants for $D$ and $D^{\ast}$, respectively.
The three-point function can be written at the hadron level
\begin{eqnarray}
\label{PHpi_DD}
&& \Pi_{\mu\nu}^{DD^{\ast}}(p, p^{\prime}, q)
\non &=&\int
d^4xd^4y \,e^{ip^{\prime} \cdot x}e^{iq \cdot y}\,\langle0|T[J^{D}(x)J^{\bar{D}^{\ast}}_{\mu}(y)J_{\nu}^{Z\dag}(0)]|0\rangle
\non &=&\frac{g_{ZDD^{\ast}}(q^2)f_{D^{\ast}}(-if_{D})f_{Z}}{(p^2-m^2_{Z}+i\epsilon)(p^{\prime 2}-m^2_{D}+i\epsilon)(q^2-m^2_{D^{\ast}}+i\epsilon)}\times
\\&&\nonumber
\left(g_{\mu\nu}-\frac{q_{\nu}p^{\prime}_{\mu}+p^{\prime}_{\mu}p^{\prime}_{\nu}}{m_Z^2}-\frac{q_{\mu}q_{\nu}}{m_{D^{\ast}}^2}+
\frac{p^{\prime}\cdot q(q_{\mu}q_{\nu}+q_{\mu}p^{\prime}_{\nu})}{m_Z^2m_{D^{\ast}}^2}\right)
\non &&+...,
\end{eqnarray}
where the coupling constant $g_{ZDD^{\ast}}(q^2)$ is defined via
\begin{eqnarray}
 \mathcal{L}=g_{ZDD^{\ast}}g^{\mu\nu}Z_{c\nu}^{+}D^-\bar{D}^{\ast 0}_{\mu}+h.c..
\end{eqnarray}

However, the evaluation of the three-point correlation function $\Pi_{\mu\nu}^{DD^{\ast}}(p, p^{\prime}, q)$
is a bit different from the other two decay modes discussed previously. Considering diagrams similar to
Fig.~\ref{Graphs}, we calculate the OPE series in the momentum spaces. As discussed above, we keep
the invariant function proportional to $1/Q^2$ in the $g_{\mu\nu}$ structure at the OPE side
\begin{eqnarray}
\nonumber
%\rho_1^{\qGqa}(s)&=&\frac{m_c\qGqb}{192\pi^2}\left(6-\frac{5m_c^2}{s}+\frac{m_c^4}{s^2}\right),
%\\ \rho_2^{\qGqa}(s)&=&-\frac{19m_c^3\qGqb}{192\pi^2(s-m_c^2)},
%\rho^{\GGa}(s)&=&\frac{\GGb}{6144\pi^4s^2}
%\left[(4s^3+19m_c^2s^2-34m_c^4s-5m_c^6)-(15m_c^2s^2-34m_c^4s-m_c^6)\log\left(1+m_c^2/Q^2\right)\right]
\rho^{\qGqa}(s)&=&\frac{m_c\qGqb}{192\pi^2}\left(6-\frac{5m_c^2}{s}+\frac{m_c^4}{s^2}-\frac{19m_c^2}{s-m_c^2}\right)
\non &&+\frac{\GGb}{6144\pi^4s^2}
\Big[(4s^3+19m_c^2s^2-34m_c^4s-5m_c^6)
\non && -(15m_c^2s^2-34m_c^4s-m_c^6)\log\left(1+\frac{m_c^2}{Q^2}\right)\Big],
\\ \label{spectral_density_DD}
\end{eqnarray}
in which $Q^2=-q^2$. Here the quark-gluon mixed condensated $\qGqb$ gives dominant contribution to the spectral density
shown above. We find that for $\qGqb$, only the color-connected diagrams contribute to the spectral density shown above, which
is different from the situations in $J/\psi\pi$ and $\eta_c\rho$ channels. For the gluon condensate $\GGb$, both the color-connected
and color-disconnected diagrams give contributions.
Assuming $p^2=p^{\prime 2}=P^2$ and performing the Borel transform ($P^2\to M_B^2$) of the three-point function,
the sum rule for the coupling constant $g_{ZDD^{\ast}}$ can be obtained as
\begin{eqnarray}
\nonumber g_{ZDD^{\ast}}(s_0, M_B^2, Q^2)&=&\frac{1}{f_{D^{\ast}}(-if_{D})f_{Z}}\frac{m^2_{Z}-m^2_{D}}{e^{-m^2_{D}/M_B^2}-e^{-m^2_{Z}/M_B^2}}
\non && \times \left(\frac{Q^2+m_{D^{\ast}}^2}{Q^2}\right)\int_{4m_c^2}^{s_0}\rho(s)e^{-s/M_B^2}ds.
\\ \label{g_DD}
\end{eqnarray}

We adopt the hadron parameters of $D^+$ and $D^{\ast0}$ from Refs.~\cite{2014-Olive-p90001-90001,2013-Dias-p16004-16004}
\begin{eqnarray}
\nonumber
m_{D^+}&=&(1869.61\pm0.10)~\mbox{MeV}, ~f_{D}=(0.18\pm0.02)~\mbox{GeV},
\\ \nonumber m_{D^{\ast0}}&=&(2006.96\pm0.10)~\mbox{MeV}, ~f_{D^{\ast}}=(0.24\pm0.02)~\mbox{GeV}.
\\ \label{parameters_DD}
\end{eqnarray}

\begin{figure}
\begin{center}
\resizebox{0.40\textwidth}{!}{%
  \includegraphics{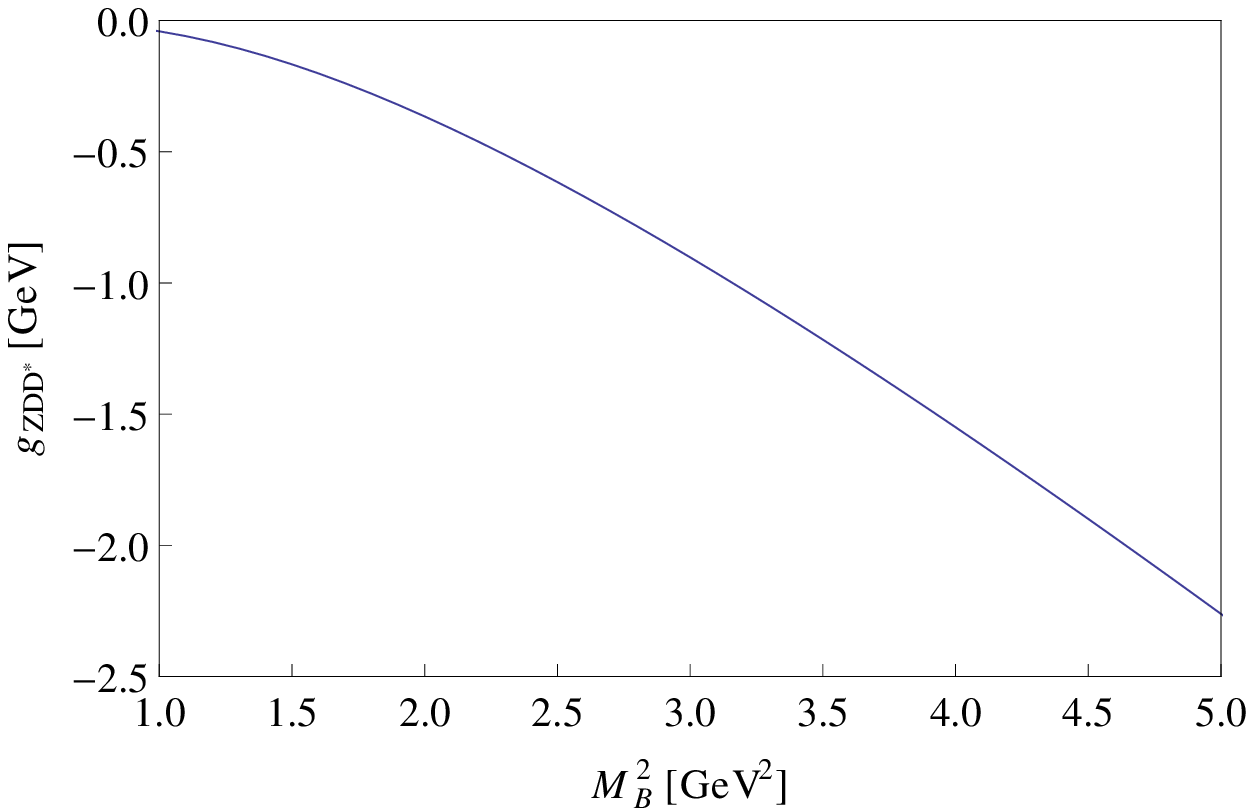}}       % Give the correct figure height in cm
\caption{Variation of the coupling constant $g_{ZDD^{\ast}}(Q^2)$ with the Borel parameter $M_B^2$ for $Q^2=20$ GeV$^2$.}
\label{fig_gDD}       % Give a unique label
\end{center}
\end{figure}

\begin{figure}
\begin{center}
\resizebox{0.40\textwidth}{!}{%
  \includegraphics{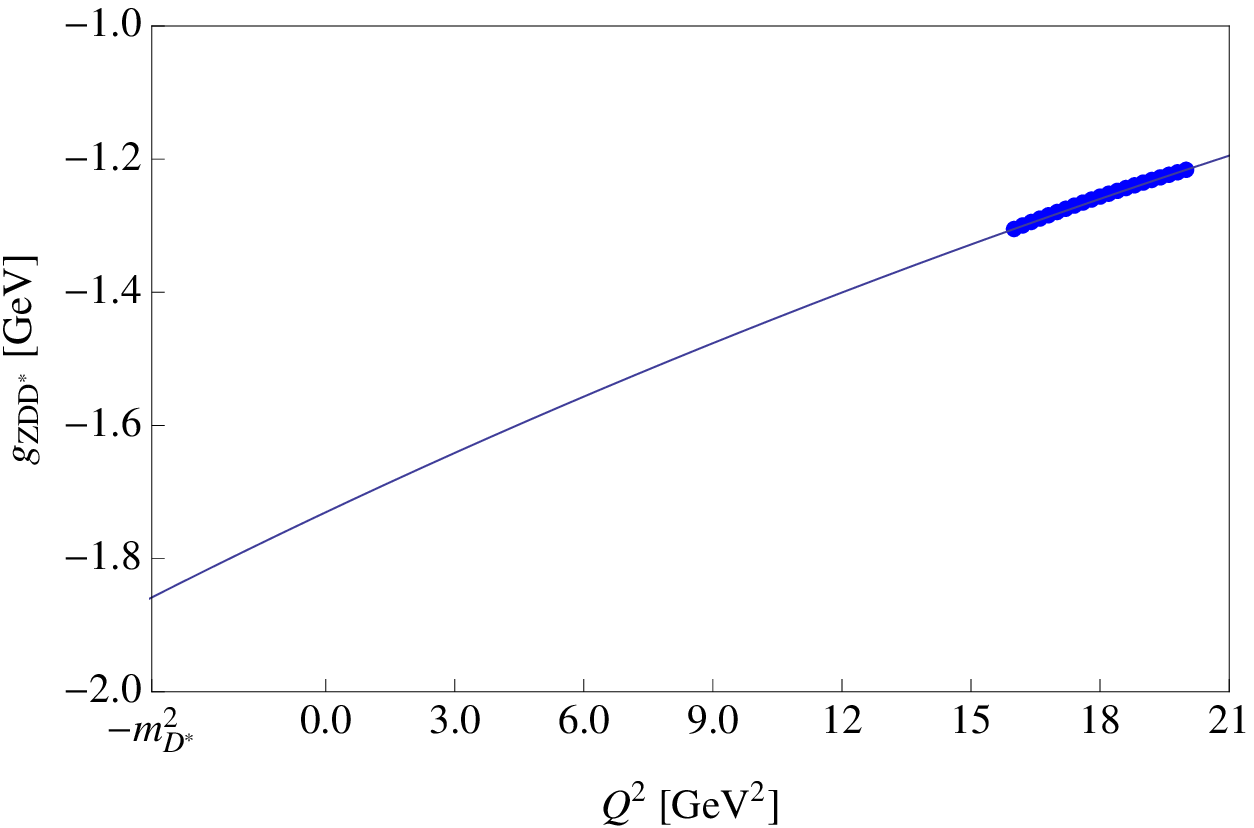}}       % Give the correct figure height in cm
\caption{Variation of the coupling constant $g_{ZDD^{\ast}}(Q^2)$ with $Q^2$. Data points shows the QCD sum-rule result with
$s_0=21$ GeV$^2$ and $M_B^2=3.4$ GeV$^2$. Solid line gives the fit of the QCD sum-rule result through Eq.~\eqref{model}
and the extrapolation of the coupling constant $g_{ZDD^{\ast}}(Q^2)$ to the physical pole $Q^2=-m_{D^{\ast}}^2$.}
\label{fig_gDD_Q}       % Give a unique label
\end{center}
\end{figure}

In Fig.~\ref{fig_gDD}, we show the variation of $g_{ZDD^{\ast}}(Q^2)$ with the Borel parameter while $Q^2=20$ GeV$^2$.
In this situation, the coupling constant increases monotonically with the momentum $Q^2$, and hence the result obtained below
should be considered as an upper bound on the comparatively small decay width  in this channel. To perform the QCD sum-rule analysis,
we adopt the Borel window $3.0\leq M_B^2\leq 3.4$ GeV$^2$ used in Ref.~\cite{2011-Chen-p34010-34010}
for the mass sum rules of the same current. We do the fitting at $M_B^2=3.4$ GeV$^2$ and $s_0=21$ GeV$^2$
in Fig.~\ref{fig_gDD_Q} using Eq.~\eqref{model} with $g_1=-1.73$ GeV and $g_2=1.77\times10^{-2}$ GeV$^2$.
With these parameters, the model Eq.~\eqref{model} can fit the QCD sum-rule result very well. To obtain the coupling
constant, we extrapolate the coupling constant $g_{ZDD^{\ast}}(Q^2)$ to the physical pole $Q^2=-m_{D^{\ast}}^2$.
Considering the uncertainties of the input parameters, the coupling constant $g_{ZDD^{\ast}}(Q^2=-m_{D^{\ast}}^2)$
is
\begin{eqnarray}
g_{ZDD^{\ast}}(Q^2=-m_{D^{\ast}}^2)=-(1.86\pm1.09)~\mbox{GeV}. \label{gvalue_DD}
\end{eqnarray}
Finally, the decay width of this decay mode can be evaluated via Eq.~\eqref{decaywidth}
\begin{eqnarray}
\Gamma(Z_c^+(4200)\to D^+\bar{D}^{\ast 0})=(6.6\pm6.4) ~\mbox{MeV}. \label{width_DD}
\end{eqnarray}

In addition, the open-charm decay mode $Z_c^+(4200)\to \bar D^0D^{\ast +}$ should also be studied.
Considering the SU(2) symmetry, the three-point function for the vertex $\\ Z_c^+(4200)\bar D^0D^{\ast +}$
is exactly the same as that for the vertex $Z_c^+(4200)D^+\bar{D}^{\ast 0}$. Using the hadron parameters of $D^0$ and $D^{\ast+}$~\cite{2014-Olive-p90001-90001}:
\begin{eqnarray}
\nonumber
m_{D^0}&=&(1864.84 \pm 0.07)~\mbox{MeV},
\non ~m_{D^{\ast+}}&=&(2010.26 \pm 0.07)~\mbox{MeV} \, ,
\end{eqnarray}
we obtain the same values of the coupling constant and the decay width for $Z_c^+(4200)\to \bar D^0D^{\ast +}$
as those for $Z_c^+(4200)\to \bar D^+D^{\ast 0}$.
%\begin{eqnarray}
%g_{ZD^0D^{\ast+}}(Q^2=-m_{D^{\ast+}}^2) &=& -(1.86\pm1.09)~\mbox{GeV} \, ,
%\\ \Gamma(Z_c^+(4200)\to D^0\bar{D}^{\ast +}) &=& (6.6\pm6.4) ~\mbox{MeV} \, ,
%\end{eqnarray}

%At last, the full decay width for the newly observed charmonium-like
%state $Z_c^+(4200)$ can be predicted as
%\begin{eqnarray}
%\Gamma_f=(353.7\pm148.9) ~\mbox{MeV},
%\end{eqnarray}
%which is consistent with the experimental value $370^{+70+70}_{-70-132}$ MeV~\cite{2014-Chilikin-p112009-112009}.

%================================================================================
%================================================================================
\section{SUMMARY}\label{sec:SUMMARY}
%================================================================================
%================================================================================
In summary, we have studied the three-point functions of the processes $Z_c(4200)^+\to J/\psi\pi^+, Z_c(4200)^+\to\eta_c\rho^+$
and $Z_c(4200)^+\to D^+\bar D^{\ast 0}$, considering $Z_c(4200)^+$ as a hidden-charm tetraquark state.
We calculate the three-point functions by including the perturbative term, quark condensate, gluon condensate
and quark-gluon mixed condensate.

To perform the QCD sum rule analysis, we expand the three-point functions in QCD side with respect to $Q^2$
and isolate the terms proportional to 1/$Q^2$ from the $g^{\mu\nu}$ tensor structure. Only the gluon condensate
and the mixed condensate give contributions to the three-point functions after this procedure. This approach is
a modification of the QCD sum-rule analysis of the $Z_c(3900)^+$ decay width
in Ref.~\cite{2013-Dias-p16004-16004}
%\end{document}
%\trem{The authors of Ref.~\cite{2013-Dias-p16004-16004}  evaluated}
where the three-point functions were evaluated by considering the color
connected (CC) diagrams associated with a different tensor structure resulting in only mixed condensate contributions.
Our predictions of the decay widths are
\begin{eqnarray}
\nonumber \Gamma({Z_c(4200)^+\to J/\psi\pi^+}) &=& (87.3\pm47.1)~{\rm MeV} \, ,
\\ \nonumber \Gamma({Z_c(4200)^+\to\eta_c\rho^+}) &=& (334.4\pm119.8)~{\rm MeV} \, ,
\\ \nonumber \Gamma({Z_c(4200)^+\to D^+\bar D^{\ast 0}}) &=& (6.6\pm6.4)~{\rm MeV} \, ,
\\ \nonumber \Gamma({Z_c(4200)^+\to \bar D^0 D^{\ast +}}) &=& (6.6\pm6.4)~{\rm MeV} \, .
\end{eqnarray}
Thus, the full decay width for $Z_c(4200)^+$ is predicted as
\begin{eqnarray}
\Gamma_{Z_c(4200)^+}=(435\pm180)~{\rm MeV} \, ,
\end{eqnarray}
which is in agreement with the experimental value of $Z_c(4200)^+$ width from
the Belle Collaboration~\cite{2014-Chilikin-p112009-112009}. It is found that the branching fraction into $J/\psi\pi$ channel is about
$24.7\%$, which is slightly suppressed compared to $71.6\%$ for the $\eta_c\rho$ channel. The study of $Z_c\to\eta_c\rho$
decay may provide useful insights on the nature of the newly observed charged $Z_c$ states, helping to discriminate
the molecule and tetraquark interpretations of the charged state family~\cite{2014-Esposito-p-}.

The study of the three-point function sum rules gives support to the tetraquark interpretation of the newly observed
$Z_c(4200)^+$ state. This conclusion is consistent with the result obtained from the mass sum rules in
Ref.~\cite{2011-Chen-p34010-34010}. The branching ratio predictions of $J/\psi\pi, \eta_c\rho, D^+\bar D^{\ast 0}$
and $\bar D^0 D^{\ast +}$ channels will be helpful for future experimental studies.

%
% BibTeX users please use
% \bibliographystyle{}
% \bibliography{}

\begin{thebibliography}{}

%\cite{Chilikin:2014bkk}
\bibitem{2014-Chilikin-p112009-112009}
  K.~Chilikin {\it et al.}  [Belle Collaboration],
  %``Observation of a new charged charmoniumlike state in $\bar{B}^0 �� J/��K^-��^+$ decays,''
  Phys.\ Rev.\ D {\bf 90}, no. 11, 112009 (2014).
  %%CITATION = ARXIV:1408.6457;%%
  %20 citations counted in INSPIRE as of 20 May 2015

%\cite{Wang:2014hta}
\bibitem{2014-Wang-p-}
  X.~L.~Wang, C.~Z.~Yuan, C.~P.~Shen, P.~Wang, A.~Abdesselam, I.~Adachi, H.~Aihara and S.~A.~Said {\it et al.},
  %``Measurement of $e^+e^- \to \pi^+\pi^-\psi(2S)$ via Initial State Radiation at Belle,''
  arXiv:1410.7641 [hep-ex].
  %%CITATION = ARXIV:1410.7641;%%
  %6 citations counted in INSPIRE as of 20 May 2015

%\cite{Choi:2007wga}
\bibitem{2008-Choi-p142001-142001}
  S.~K.~Choi {\it et al.}  [Belle Collaboration],
  %``Observation of a resonance-like structure in the pi+- psi-prime mass distribution in exclusive B ---> K pi+- psi-prime decays,''
  Phys.\ Rev.\ Lett.\  {\bf 100}, 142001 (2008).
  %%CITATION = ARXIV:0708.1790;%%
  %376 citations counted in INSPIRE as of 20 May 2015

%\cite{Aaij:2014jqa}
\bibitem{2014-Aaij-p222002-222002}
  R.~Aaij {\it et al.}  [LHCb Collaboration],
  %``Observation of the resonant character of the $Z(4430)^-$ state,''
  Phys.\ Rev.\ Lett.\  {\bf 112}, no. 22, 222002 (2014).
  %%CITATION = ARXIV:1404.1903;%%
  %71 citations counted in INSPIRE as of 20 May 2015

%\cite{Mizuk:2008me}
\bibitem{2008-Mizuk-p72004-72004}
  R.~Mizuk {\it et al.}  [Belle Collaboration],
  %``Observation of two resonance-like structures in the pi+ chi(c1) mass distribution in exclusive anti-B0 ---> K- pi+ chi(c1) decays,''
  Phys.\ Rev.\ D {\bf 78}, 072004 (2008).
  %%CITATION = ARXIV:0806.4098;%%
  %201 citations counted in INSPIRE as of 20 May 2015

%\cite{Ablikim:2013mio}
\bibitem{2013-Ablikim-p252001-252001}
  M.~Ablikim {\it et al.}  [BESIII Collaboration],
  Phys.\ Rev.\ Lett.\  {\bf 110}, 252001 (2013).

%\cite{Liu:2013dau}
\bibitem{2013-Liu-p252002-252002}
  Z.~Q.~Liu {\it et al.}  [Belle Collaboration],
  %``Study of $e^+e^- ��^+ ��^- J/��$ and Observation of a Charged Charmoniumlike State at Belle,''
  Phys.\ Rev.\ Lett.\  {\bf 110}, 252002 (2013).
  %%CITATION = ARXIV:1304.0121;%%
  %204 citations counted in INSPIRE as of 20 May 2015

%\cite{Xiao:2013iha}
\bibitem{2013-Xiao-p366-370}
  T.~Xiao, S.~Dobbs, A.~Tomaradze and K.~K.~Seth,
  %``Observation of the Charged Hadron $Z_c^{\pm}(3900)$ and Evidence for the Neutral $Z_c^0(3900)$ in $e^+e^-\to \pi\pi J/\psi$ at $\sqrt{s}=4170$ MeV,''
  Phys.\ Lett.\ B {\bf 727}, 366 (2013).
  %%CITATION = ARXIV:1304.3036;%%
  %130 citations counted in INSPIRE as of 20 May 2015

%\cite{Ablikim:2013emm}
\bibitem{2014-Ablikim-p132001-132001}
  M.~Ablikim {\it et al.}  [BESIII Collaboration],
  %``Observation of a charged charmoniumlike structure in $e^+e^- \to (D^{*} \bar{D}^{*})^{\pm} \pi^\mp$ at $\sqrt{s}=4.26$GeV,''
  Phys.\ Rev.\ Lett.\  {\bf 112}, no. 13, 132001 (2014).
  %%CITATION = ARXIV:1308.2760;%%
  %112 citations counted in INSPIRE as of 20 May 2015

%\cite{Ablikim:2013wzq}
\bibitem{2013-Ablikim-p242001-242001}
  M.~Ablikim {\it et al.}  [BESIII Collaboration],
  %``Observation of a Charged Charmoniumlike Structure $Z_c$(4020) and Search for the $Z_c$(3900) in $e^+e^- \to ��^+��^-h_c$,''
  Phys.\ Rev.\ Lett.\  {\bf 111}, no. 24, 242001 (2013).
  %%CITATION = ARXIV:1309.1896;%%
  %107 citations counted in INSPIRE as of 20 May 2015

%\cite{Collaboration:2011gja}
\bibitem{2011-Adachi-p-}
  I.~Adachi [Belle Collaboration],
  %``Observation of two charged bottomonium-like resonances,''
  arXiv:1105.4583 [hep-ex].
  %%CITATION = ARXIV:1105.4583;%%
  %129 citations counted in INSPIRE as of 20 May 2015


%\cite{Close:2003sg}
\bibitem{Close:2003sg}
  F.~E.~Close and P.~R.~Page,
  %``The D*0 anti-D0 threshold resonance,''
  Phys.\ Lett.\ B {\bf 578}, 119 (2004).
  %%CITATION = HEP-PH/0309253;%%
  %314 citations counted in INSPIRE as of 20 May 2015

%\cite{Braaten:2003he}
\bibitem{Braaten:2003he}
  E.~Braaten and M.~Kusunoki,
  %``Low-energy universality and the new charmonium resonance at 3870-MeV,''
  Phys.\ Rev.\ D {\bf 69}, 074005 (2004).
  %%CITATION = HEP-PH/0311147;%%
  %173 citations counted in INSPIRE as of 20 May 2015

%\cite{Tornqvist:2004qy}
\bibitem{Tornqvist:2004qy}
  N.~A.~Tornqvist,
  %``Isospin breaking of the narrow charmonium state of Belle at 3872-MeV as a deuson,''
  Phys.\ Lett.\ B {\bf 590}, 209 (2004).
  %%CITATION = HEP-PH/0402237;%%
  %386 citations counted in INSPIRE as of 20 May 2015

%\cite{Swanson:2006st}
\bibitem{Swanson:2006st}
  E.~S.~Swanson,
  %``The New heavy mesons: A Status report,''
  Phys.\ Rept.\  {\bf 429}, 243 (2006).
  %%CITATION = HEP-PH/0601110;%%
  %525 citations counted in INSPIRE as of 20 May 2015

%\cite{Fleming:2007rp}
\bibitem{Fleming:2007rp}
  S.~Fleming, M.~Kusunoki, T.~Mehen and U.~van Kolck,
  %``Pion interactions in the $X(3872)$,''
  Phys.\ Rev.\ D {\bf 76}, 034006 (2007).
  %%CITATION = HEP-PH/0703168;%%
  %92 citations counted in INSPIRE as of 20 May 2015

%\cite{Braaten:2007dw}
\bibitem{Braaten:2007dw}
  E.~Braaten and M.~Lu,
  %``Line shapes of the X(3872),''
  Phys.\ Rev.\ D {\bf 76}, 094028 (2007).
  %%CITATION = ARXIV:0709.2697;%%
  %85 citations counted in INSPIRE as of 20 May 2015

%\cite{Guo:2013sya}
\bibitem{Guo:2013sya}
  F.~K.~Guo, C.~Hidalgo-Duque, J.~Nieves and M.~P.~Valderrama,
  %``Consequences of Heavy Quark Symmetries for Hadronic Molecules,''
  Phys.\ Rev.\ D {\bf 88}, 054007 (2013).
  %%CITATION = ARXIV:1303.6608;%%
  %72 citations counted in INSPIRE as of 20 May 2015

%\cite{Maiani:2004vq}
\bibitem{Maiani:2004vq}
  L.~Maiani, F.~Piccinini, A.~D.~Polosa and V.~Riquer,
  %``Diquark-antidiquarks with hidden or open charm and the nature of X(3872),''
  Phys.\ Rev.\ D {\bf 71}, 014028 (2005).
  %%CITATION = HEP-PH/0412098;%%
  %431 citations counted in INSPIRE as of 20 May 2015

%\cite{Ali:2011ug}
\bibitem{Ali:2011ug}
  A.~Ali, C.~Hambrock and W.~Wang,
  %``Tetraquark Interpretation of the Charged Bottomonium-like states $Z_b^{+-}(10610)$ and $Z_b^{+-}(10650)$ and Implications,''
  Phys.\ Rev.\ D {\bf 85}, 054011 (2012).
  %%CITATION = ARXIV:1110.1333;%%
  %47 citations counted in INSPIRE as of 20 May 2015

%\cite{Ebert:2008kb}
\bibitem{Ebert:2008kb}
  D.~Ebert, R.~N.~Faustov and V.~O.~Galkin,
  %``Excited heavy tetraquarks with hidden charm,''
  Eur.\ Phys.\ J.\ C {\bf 58}, 399 (2008).
  %%CITATION = ARXIV:0808.3912;%%
  %55 citations counted in INSPIRE as of 20 May 2015

%\cite{Dubynskiy:2008mq}
\bibitem{Dubynskiy:2008mq}
  S.~Dubynskiy and M.~B.~Voloshin,
  %``Hadro-Charmonium,''
  Phys.\ Lett.\ B {\bf 666}, 344 (2008).
  %%CITATION = ARXIV:0803.2224;%%
  %82 citations counted in INSPIRE as of 20 May 2015

%\cite{Danilkin:2011sh}
\bibitem{Danilkin:2011sh}
  I.~V.~Danilkin, V.~D.~Orlovsky and Y.~A.~Simonov,
  %``Hadron interaction with heavy quarkonia,''
  Phys.\ Rev.\ D {\bf 85}, 034012 (2012).
  %%CITATION = ARXIV:1106.1552;%%
  %29 citations counted in INSPIRE as of 20 May 2015

%\cite{Matheus:2009vq}
\bibitem{Matheus:2009vq}
  R.~D.~Matheus, F.~S.~Navarra, M.~Nielsen and C.~M.~Zanetti,
  %``QCD Sum Rules for the X(3872) as a mixed molecule-charmoniun state,''
  Phys.\ Rev.\ D {\bf 80}, 056002 (2009).
  %%CITATION = ARXIV:0907.2683;%%
  %51 citations counted in INSPIRE as of 20 May 2015

%\cite{Liu:2007bf}
\bibitem{2008-Liu-p34003-34003}
  X.~Liu, Y.~R.~Liu, W.~Z.~Deng and S.~L.~Zhu,
  %``Is Z+(4430) a loosely bound molecular state?,''
  Phys.\ Rev.\ D {\bf 77}, 034003 (2008).
  %%CITATION = ARXIV:0711.0494;%%
  %87 citations counted in INSPIRE as of 20 May 2015

%\cite{Lee:2007gs}
\bibitem{2008-Lee-p28-32}
  S.~H.~Lee, A.~Mihara, F.~S.~Navarra and M.~Nielsen,
  %``QCD sum rules study of the meson Z+(4430),''
  Phys.\ Lett.\ B {\bf 661}, 28 (2008).
  %%CITATION = ARXIV:0710.1029;%%
  %72 citations counted in INSPIRE as of 20 May 2015

%\cite{Meng:2007fu}
\bibitem{2007-Meng-p-}
  C.~Meng and K.~T.~Chao,
  %``Z+(4430) as a resonance in the D(1)(D(1)-prime)D* channel,''
  arXiv:0708.4222 [hep-ph].
  %%CITATION = ARXIV:0708.4222;%%
  %82 citations counted in INSPIRE as of 20 May 2015

%\cite{Ding:2007ar}
\bibitem{2007-Ding-p-}
  G.~J.~Ding,
  %``Understanding the Charged Meson Z(4430),''
  arXiv:0711.1485 [hep-ph].
  %%CITATION = ARXIV:0711.1485;%%
  %39 citations counted in INSPIRE as of 20 May 2015

%\cite{Bracco:2008jj}
\bibitem{2009-Bracco-p240-244}
  M.~E.~Bracco, S.~H.~Lee, M.~Nielsen and R.~Rodrigues da Silva,
  %``The Meson Z+(4430) as a tetraquark state,''
  Phys.\ Lett.\ B {\bf 671}, 240 (2009).
  %%CITATION = ARXIV:0807.3275;%%
  %46 citations counted in INSPIRE as of 20 May 2015

%\cite{Maiani:2008zz}
\bibitem{2008-Maiani-p73004-73004}
  L.~Maiani, A.~D.~Polosa and V.~Riquer,
  %``The charged Z(4430) in the diquark-antidiquark picture,''
  New J.\ Phys.\  {\bf 10}, 073004 (2008).
  %%CITATION = NJOPF,10,073004;%%
  %24 citations counted in INSPIRE as of 20 May 2015

%\cite{Maiani:2014aja}
\bibitem{2014-Maiani-p114010-114010}
  L.~Maiani, F.~Piccinini, A.~D.~Polosa and V.~Riquer,
  %``The Z(4430) and a New Paradigm for Spin Interactions in Tetraquarks,''
  Phys.\ Rev.\ D {\bf 89}, no. 11, 114010 (2014).
  %%CITATION = ARXIV:1405.1551;%%
  %32 citations counted in INSPIRE as of 20 May 2015

%\cite{Wang:2013cya}
\bibitem{2013-Wang-p132003-132003}
  Q.~Wang, C.~Hanhart and Q.~Zhao,
  %``Decoding the riddle of $Y(4260)$ and $Z_c(3900)$,''
  Phys.\ Rev.\ Lett.\  {\bf 111}, no. 13, 132003 (2013).
  %%CITATION = ARXIV:1303.6355;%%
  %92 citations counted in INSPIRE as of 20 May 2015

%\cite{Aceti:2014uea}
\bibitem{2014-Aceti-p16003-16003}
  F.~Aceti, M.~Bayar, E.~Oset, A.~Martinez Torres, K.~P.~Khemchandani, J.~M.~Dias, F.~S.~Navarra and M.~Nielsen,
  %``Prediction of an $I=1$ $D \bar D^*$ state and relationship to the claimed $Z_c(3900)$, $Z_c(3885)$,''
  Phys.\ Rev.\ D {\bf 90}, no. 1, 016003 (2014).
  %%CITATION = ARXIV:1401.8216;%%
  %12 citations counted in INSPIRE as of 20 May 2015

%\cite{Zhao:2014gqa}
\bibitem{2014-Zhao-p94026-94026}
  L.~Zhao, L.~Ma and S.~L.~Zhu,
  %``Spin-orbit force, recoil corrections, and possible $B \bar{B}^{*}$ and $D \bar{D}^{*}$  molecular states,''
  Phys.\ Rev.\ D {\bf 89}, no. 9, 094026 (2014).
  %%CITATION = ARXIV:1403.4043;%%
  %5 citations counted in INSPIRE as of 20 May 2015

%\cite{He:2013nwa}
\bibitem{2013-He-p2635-2635}
  J.~He, X.~Liu, Z.~F.~Sun and S.~L.~Zhu,
  %``$Z_c(4025)$ as the hadronic molecule with hidden charm,''
  Eur.\ Phys.\ J.\ C {\bf 73}, no. 11, 2635 (2013).
  %%CITATION = ARXIV:1308.2999;%%
  %25 citations counted in INSPIRE as of 20 May 2015

%\cite{Liu:2008tn}
\bibitem{2009-Liu-p411-428}
  X.~Liu, Z.~G.~Luo, Y.~R.~Liu and S.~L.~Zhu,
  %``X(3872) and Other Possible Heavy Molecular States,''
  Eur.\ Phys.\ J.\ C {\bf 61}, 411 (2009).
  %%CITATION = ARXIV:0808.0073;%%
  %121 citations counted in INSPIRE as of 20 May 2015

%\cite{Agashe:2014kda}
\bibitem{2014-Olive-p90001-90001}
  K.~A.~Olive {\it et al.}  [Particle Data Group Collaboration],
  %``Review of Particle Physics,''
  Chin.\ Phys.\ C {\bf 38}, 090001 (2014).
  %%CITATION = CHPHD,C38,090001;%%
  %1078 citations counted in INSPIRE as of 20 May 2015

%\cite{Brambilla:2010cs}
\bibitem{Brambilla:2010cs}
  N.~Brambilla, S.~Eidelman, B.~K.~Heltsley, R.~Vogt, G.~T.~Bodwin, E.~Eichten, A.~D.~Frawley and A.~B.~Meyer {\it et al.},
  %``Heavy quarkonium: progress, puzzles, and opportunities,''
  Eur.\ Phys.\ J.\ C {\bf 71}, 1534 (2011).
  %%CITATION = ARXIV:1010.5827;%%
  %751 citations counted in INSPIRE as of 20 May 2015

%\cite{Nielsen:2009uh}
\bibitem{Nielsen:2009uh}
  M.~Nielsen, F.~S.~Navarra and S.~H.~Lee,
  %``New Charmonium States in QCD Sum Rules: A Concise Review,''
  Phys.\ Rept.\  {\bf 497}, 41 (2010).
  %%CITATION = ARXIV:0911.1958;%%
  %131 citations counted in INSPIRE as of 20 May 2015

%\cite{Esposito:2014rxa}
\bibitem{Esposito:2014rxa}
  A.~Esposito, A.~L.~Guerrieri, F.~Piccinini, A.~Pilloni and A.~D.~Polosa,
  %``Four-Quark Hadrons: an Updated Review,''
  Int.\ J.\ Mod.\ Phys.\ A {\bf 30}, no. 04n05, 1530002 (2014).
  %%CITATION = ARXIV:1411.5997;%%
  %17 citations counted in INSPIRE as of 20 May 2015

%\cite{Olsen:2014qna}
\bibitem{Olsen:2014qna}
  S.~L.~Olsen,
  %``A New Hadron Spectroscopy,''
  Front.\ Phys.\ China.\  {\bf 10}, 121 (2015).
  %%CITATION = ARXIV:1411.7738;%%
  %4 citations counted in INSPIRE as of 20 May 2015

%\cite{Liu:2013waa}
\bibitem{Liu:2013waa}
  X.~Liu,
  %``An overview of $XYZ$ new particles,''
  Chin.\ Sci.\ Bull.\  {\bf 59}, 3815 (2014).
  %%CITATION = ARXIV:1312.7408;%%
  %22 citations counted in INSPIRE as of 20 May 2015

%\cite{Zhao:2014qva}
\bibitem{2014-Zhao-p-}
  L.~Zhao, W.~Z.~Deng and S.~L.~Zhu,
  %``Hidden-Charm Tetraquarks and Charged $Z_c$ States,''
  Phys.\ Rev.\ D {\bf 90}, no. 9, 094031 (2014).
  %%CITATION = ARXIV:1408.3924;%%
  %7 citations counted in INSPIRE as of 20 May 2015

%\cite{Prelovsek:2014swa}
\bibitem{2014-Prelovsek-p-}
  S.~Prelovsek, C.~B.~Lang, L.~Leskovec and D.~Mohler,
  %``Study of the $Z_c^+$ channel using lattice QCD,''
  Phys.\ Rev.\ D {\bf 91}, no. 1, 014504 (2015).
  %%CITATION = ARXIV:1405.7623;%%
  %16 citations counted in INSPIRE as of 20 May 2015

%\cite{Qiao:2013raa}
\bibitem{Qiao:2013raa}
  C.~F.~Qiao and L.~Tang,
  %``Estimating the mass of the hidden charm $1^+(1^{+})$ tetraquark state via QCD sum rules,''
  Eur.\ Phys.\ J.\ C {\bf 74}, no. 10, 3122 (2014).
  %%CITATION = ARXIV:1307.6654;%%
  %8 citations counted in INSPIRE as of 20 May 2015

%\cite{Cui:2013vfa}
\bibitem{2013-Cui-p-}
  C.~Y.~Cui, Y.~L.~Liu and M.~Q.~Huang,
  %``Could $Z_{c}(4025)$ be a $J^{P}=1^{+}$ $D^{*}\bar{D^{*}}$ molecular state?,''
  arXiv:1308.3625 [hep-ph].
  %%CITATION = ARXIV:1308.3625;%%
  %11 citations counted in INSPIRE as of 20 May 2015

%\cite{Dias:2013xfa}
\bibitem{2013-Dias-p16004-16004}
  J.~M.~Dias, F.~S.~Navarra, M.~Nielsen and C.~M.~Zanetti,
  %``$Z^+_c$(3900) decay width in QCD sum rules,''
  Phys.\ Rev.\ D {\bf 88}, no. 1, 016004 (2013).
  %%CITATION = ARXIV:1304.6433;%%
  %28 citations counted in INSPIRE as of 20 May 2015

%\cite{Narison:2010pd}
\bibitem{Narison:2010pd}
  S.~Narison, F.~S.~Navarra and M.~Nielsen,
  %``On the nature of the X(3872) from QCD,''
  Phys.\ Rev.\ D {\bf 83}, 016004 (2011).
  %%CITATION = ARXIV:1006.4802;%%
  %21 citations counted in INSPIRE as of 20 May 2015

%\cite{Zhang:2013aoa}
\bibitem{Zhang:2013aoa}
  J.~R.~Zhang,
  %``Improved QCD sum rule study of $Z_{c}(3900)$ as a $\bar{D}D^{*}$ molecular state,''
  Phys.\ Rev.\ D {\bf 87}, no. 11, 116004 (2013).
  %%CITATION = ARXIV:1304.5748;%%
  %28 citations counted in INSPIRE as of 20 May 2015

%\cite{Wang:2014gwa}
\bibitem{Wang:2014gwa}
  Z.~G.~Wang,
  %``Reanalysis of the $Y(3940)$, $Y(4140)$, $Z_c(4020)$, $Z_c(4025)$ and $Z_b(10650)$ as molecular states with QCD sum rules,''
  Eur.\ Phys.\ J.\ C {\bf 74}, no. 7, 2963 (2014).
  %%CITATION = ARXIV:1403.0810;%%
  %8 citations counted in INSPIRE as of 20 May 2015

%\cite{Chen:2010ze}
\bibitem{2011-Chen-p34010-34010}
  W.~Chen and S.~L.~Zhu,
  %``The Vector and Axial-Vector Charmonium-like States,''
  Phys.\ Rev.\ D {\bf 83}, 034010 (2011).
  %%CITATION = ARXIV:1010.3397;%%
  %41 citations counted in INSPIRE as of 20 May 2015

%\cite{Chen:2012pe}
\bibitem{2012-Chen-p1003-1003}
  W.~Chen and S.~L.~Zhu,
  %``Spin-1 charmonium-like states in QCD sum rule,''
  EPJ Web Conf.\  {\bf 20}, 01003 (2012).
  %%CITATION = ARXIV:1209.4748;%%
  %4 citations counted in INSPIRE as of 20 May 2015

%\cite{Shifman:1978bx}
\bibitem{1979-Shifman-p385-447}
  M.~A.~Shifman, A.~I.~Vainshtein and V.~I.~Zakharov,
  %``QCD and Resonance Physics. Sum Rules,''
  Nucl.\ Phys.\ B {\bf 147}, 385 (1979).
  %%CITATION = NUPHA,B147,385;%%
  %4425 citations counted in INSPIRE as of 20 May 2015

%\cite{Reinders:1984sr}
\bibitem{1985-Reinders-p1-1}
  L.~J.~Reinders, H.~Rubinstein and S.~Yazaki,
  %``Hadron Properties from QCD Sum Rules,''
  Phys.\ Rept.\  {\bf 127}, 1 (1985).
  %%CITATION = PRPLC,127,1;%%
  %1315 citations counted in INSPIRE as of 20 May 2015

%\cite{Colangelo:2000dp}
\bibitem{2000-Colangelo-p1495-1576}
  P.~Colangelo and A.~Khodjamirian,
  %``QCD sum rules, a modern perspective,''
  In *Shifman, M. (ed.): At the frontier of particle physics, vol. 3* 1495-1576.
  %%CITATION = HEP-PH/0010175;%%
  %416 citations counted in INSPIRE as of 20 May 2015

%\cite{Ioffe:1983ju}
\bibitem{Ioffe-1983ju}
  B.~L.~Ioffe and A.~V.~Smilga,
  %``Nucleon Magnetic Moments and Magnetic Properties of Vacuum in QCD,''
  Nucl.\ Phys.\ B {\bf 232}, 109 (1984).
  %%CITATION = NUPHA,B232,109;%%
  %390 citations counted in INSPIRE as of 21 May 2015

%\cite{Eidemuller:2005jm}
\bibitem{Eidemuller-2005jm}
  M.~Eidemuller, F.~S.~Navarra, M.~Nielsen and R.~Rodrigues da Silva,
  %``Pentaquark decay width in QCD sum rules,''
  Phys.\ Rev.\ D {\bf 72}, 034003 (2005).
  %%CITATION = HEP-PH/0503193;%%
  %15 citations counted in INSPIRE as of 21 May 2015

%\cite{Chen:2014fza}
\bibitem{2014-Chen-p13-40}
  W.~Chen, T.~G.~Steele and S.~L.~Zhu,
  %``Heavy tetraquark states and quarkonium hybrids,''
  Universe {\bf 2}, 13 (2014).
  %%CITATION = ARXIV:1403.7457;%%
  %3 citations counted in INSPIRE as of 20 May 2015

%\cite{Yang:1993bp}
\bibitem{1993-Yang-p3001-3012}
  K.~C.~Yang, W.~Y.~P.~Hwang, E.~M.~Henley and L.~S.~Kisslinger,
  %``QCD sum rules and neutron proton mass difference,''
  Phys.\ Rev.\ D {\bf 47}, 3001 (1993).
  %%CITATION = PHRVA,D47,3001;%%
  %78 citations counted in INSPIRE as of 20 May 2015

%\cite{Narison:2011rn}
\bibitem{2012-Narison-p259-263}
  S.~Narison,
  %``Gluon Condensates and m_b(m_b) from QCD-Exponential Moments at Higher Orders,''
  Phys.\ Lett.\ B {\bf 707}, 259 (2012).
  %%CITATION = ARXIV:1105.5070;%%
  %40 citations counted in INSPIRE as of 20 May 2015

%\cite{Narison:1997ma}
\bibitem{1997-Narison-p238-243}
  S.~Narison,
  %``QCD '96. Proceedings, 4th Conference, Montpellier, France, July 4-12, 1996,''
  Nucl.\ Phys.\ Proc.\ Suppl.\  {\bf 54A} (1997).
  %%CITATION = NUPHZ,54A,;%%

%\cite{Faccini:2013lda}
\bibitem{2013-Maiani-p111102-111102}
  L.~Maiani, V.~Riquer, R.~Faccini, F.~Piccinini, A.~Pilloni and A.~D.~Polosa,
  %``A $J^{PG}=1^{++}$ Charged Resonance in the $Y(4260) \to \pi^+ \pi^- J/\psi$ Decay?,''
  Phys.\ Rev.\ D {\bf 87}, no. 11, 111102 (2013).
  %%CITATION = ARXIV:1303.6857;%%
  %59 citations counted in INSPIRE as of 20 May 2015

%\cite{Esposito:2014hsa}
\bibitem{2014-Esposito-p-}
  A.~Esposito, A.~L.~Guerrieri and A.~Pilloni,
  %``Probing the nature of Z_c states via the eta_c rho decay,''
  Phys.\ Lett.\ B {\bf 746}, 194 (2015).
  %%CITATION = ARXIV:1409.3551;%%
  %3 citations counted in INSPIRE as of 20 May 2015

\end{thebibliography}

\end{document}